\documentclass{aa}
\usepackage{graphicx}
\usepackage{hyperref}
\usepackage{xcolor}
\usepackage{txfonts}

\begin{document}

   \title{SN~2023adsy: A normal type Ia Supernova at z=2.9 }

   \author{ J. Vink{\'o}\inst{1, 2, 3} \and
            E. Reg{\H o}s\inst{1}
          }
          
   \institute{HUN-REN CSFK Konkoly Observatory, MTA Centre of Excellence, Konkoly Thege Mikl\'os {\'u}t 15-17., 1121 Budapest, Hungary
   \and
   Department of Experimental Physics, Institute of Physics, University of Szeged, D{\'o}m t{\'e}r 9, 6720 Szeged, Hungary \\ 
              \email{vinko@konkoly.hu}
        \and 
        ELTE E{\"o}tv{\"o}s Lor{\'a}nd University, Institute of Physics and Astronomy, P{\'a}zm{\'a}ny P{\'e}ter s{\'e}t{\'a}ny 1/A, 1117 Budapest, Hungary \\ 
         }

   \date{Received February 28 2025}
 
  \abstract
   {Supernovae (SNe) discovered in high-redshift ($z > 2$) galaxies by the \textit{James Webb} Space Telescope (JWST) provide a unique opportunity to extend the Hubble diagram beyond $z \sim 1.5$ and constrain the cosmic star formation rate in the early Universe. 
   }
   {SN~2023adsy, a type Ia supernova discovered by JWST at $z=2.9$, was found to be a peculiar event, since it is extremely red and faint, but it shows a very similar rest-frame light-curve decline rate to the majority of low-redshift SNe Ia. We examine whether the red color and faint peak magnitude could also be explained by significant reddening and extinction due to dust within the host galaxy. }
   {We used two light-curve fitting codes, SALT3-NIR and BayeSN, which have templates extended to the near-infrared, to re-fit the published NIRCam photometry, assuming a ``normal'' SN Ia and Milky-Way dust extinction law. We also examined NIRCam photometry of the host galaxy taken before the SN discovery by comparing its spectral energy distribution with galaxy templates.}
   {The NIRCam photometry can be fit reasonably well with a slowly declining but otherwise ``normal'' SN Ia template that suffers significant reddening ($E(B-V)_{\rm host} \gtrsim 0.5$ mag). Photometry of the host galaxy suggests a blue, star-forming galaxy, where the presence of a significant amount of dust cannot be ruled out. }
   {A comparison of the inferred luminosity distance with the prediction of the $\Lambda$CDM cosmology on the Hubble diagram suggests no significant evolution of the SN Ia peak luminosity at $z>2$ redshifts. We also show that the discovery of a single SN Ia between $2 < z < 3$ within the area of the JADES survey during approximately one year is consistent with the current estimates for the SN Ia rates at such redshifts. }

   \keywords{ (Stars:) supernovae: general; (Stars:) supernovae: individual: SN~2023adsy; (Cosmology:) distance scale; 
               }

   \maketitle

\section{Introduction} \label{sec:intro}

Thermonuclear (type Ia) supernovae (SNe) discovered at high ($z > 2$) redshifts offer a great opportunity for testing both the cosmic star formation rate \citep[SFR,][]{takeshi05, rv19, yan23} as well as the nature of dark energy via cosmological models \citep{lu22}. Recently, SN~2023adsy was discovered by \citet[][P24 hereafter]{pierel24} as a type Ia SN at $z = 2.9$ during the JWST Advanced Deep Extragalactic Survey (JADES) \citep{jades23}. Interestingly, P24 find that SN~2023adsy looks like a peculiar SN Ia: its light curve and spectrum suggest a normal type Ia, but it has a very red color reminiscent of an extremely low-luminosity SN Ia (its absolute peak brightness in the B band was found to be $M_B \sim -16.45$ mag, about 3 mag lower than that of normal SNe Ia), despite its normal light-curve decline rate and relatively high Ca~II expansion velocity ($\sim 18,000$ km~s$^{-1}$). P24 also point out that with such a faint peak brightness and red color, SN~2023adsy could also be a Ca-rich SN Ia, even though it is $\sim 1$ mag brighter than the typical Ca-rich event SN~2016hnk. They note that its peak brightness is more similar to that of a 91bg-like SNe Ia, but the relatively normal decline rate of its light curve is incompatible with a fast-declining 91bg-like SN. Finally, P24 find that the standardized peak absolute magnitude of SN~2023adsy is consistent with the prediction of the $\Lambda$CDM cosmological model assuming $H_0 = 70$ km~s$^{-1}$~Mpc$^{-1}$ and $\Omega_{\rm M} = 0.315$, despite its peculiarities mentioned above. 

P24 argued that the spectral energy distribution (SED) fitting of the host galaxy of SN~2023adsy (JADES-GS+53.13485$-$27.82088) indicates a low-mass ($\sim 10^8$~M$_\odot$), low-metallicity ($\sim 0.3 Z_\odot$), low-extinction ($A_V < 0.1$ mag) galaxy, which suggests that the red color of the SN is probably not due to dust extinction. However, if SN~2023adsy were intrinsically so red as observed, it would be a very peculiar SN Ia, as mentioned above. Since SN~2023adsy is the first spectroscopically confirmed SN Ia beyond $z > 2$, its peculiar nature might imply additional possibilities about the physical differences between high- and low-redshift SNe Ia. 

In this paper, we revisit the issue of the red color of SN~2023adsy by investigating whether it could be simply due to dust extinction within its host galaxy, even though the galaxy SED shows relatively low dust extinction. We reanalyze the \textit{James Webb} Space Telescope (JWST) NIRCam photometry of SN~2023adsy published by P24, and examine how the parameters of the best-fit light-curve templates change if host extinction is also allowed in the model. The data and the fitting methodology is explained in Sect.~\ref{sec:model}, while the results are presented and discussed in Sect.~\ref{sec:results}.   

\section{Data and modeling}\label{sec:model}

We used JWST NIRCam photometry of SN~2023adsy from P24, taken with the F150W, F200W, F277W, F356W, and F444W filters. Data taken with F090W and F115W are not considered, as P24 only obtain upper limits for SN~2023adsy using those filters. 

\subsection{Fitting with SALT3-NIR}\label{sec:salt3-nir} 

First, we fit the NIRCam photometry using the SALT3-NIR light-curve model \citep{pierel22} that extends the spectral coverage of the SALT3 templates \citep{kenworthy21} into the near-IR, up to $\sim 2$ microns. SALT3-NIR contains SED flux templates that are fitted to the observations using the following parameters:
\begin{itemize}
\item{$x_0$: the flux-scaling factor (``amplitude'') of the best-fit template};
\item{$x_1$: the time-scaling factor (``stretch'') of the best-fit template};
\item{$c$: the color coefficient.}
\end{itemize}
Since $x_0$ is very low ($\sim 10^{-7}$) for $z > 2$ SNe, the rest-frame $B$-band peak magnitude, $m_B$, can also be used instead of $x_0$. From the Pantheon+SH0ES sample \citep{riess22}, we found that $m_B$ is related to $x_0$ via
$m_B = -2.5 \log_{10}(x_0) + 10.607$. 

The peak absolute magnitude of the SN in the rest-frame $B$ band is calculated via the Tripp-relation \citep{tripp98} as
\begin{equation}
M_B = M_B^0 - \alpha x_1 + \beta c,
\label{eq:Mbpeak}
\end{equation}
where $M_B^0$ is the standardized peak absolute magnitude of the SN in the $B$ band, $x_1$ and $c$ are the best-fit SALT3 stretch and color parameters. In Equation~\ref{eq:Mbpeak} $\alpha$ and $\beta$ are the nuisance parameters that minimize the dispersion of the SNe Ia measurements on the Hubble diagram. 

In this study, we adopt $M_B^0 = -19.21$ \citep{riess22}, $\alpha = 0.148$, and $\beta = 3.09$ \citep{brout22}, which are based on the analysis of the Pantheon+ and SH0ES samples of SNe Ia. Using the data from \citet{riess22}, we found the value of $M_B^0$  by shifting the SNe Ia distance scale to a common zero point determined by other distance calibrators (Cepheids). 
Finally, from the best-fit SALT3-NIR parameters, the distance modulus of the SN was given by
\begin{equation}
\mu = m_B - M_B = m_B - M_B^0 + \alpha x_1 - \beta c.
\label{eq:mu}
\end{equation}

Note that in Eq.~(\ref{eq:mu}) we ignore the bias-correction term $\delta_B$ \citep{kessler17}, since SN~2023adsy turns out to be redder than the majority of the normal SNe Ia (see Sect.~\ref{sec:results}), while $\delta_B$ is more important for bluer SNe Ia \citep{kessler17}.  P24 also find that, for SN~2023adsy, $\delta_B$ is negligibly small. 

By definition, the SALT3 model does not contain information on extinction from dust either in the Milky Way or within the host galaxy. In the NIR, the Milky Way extinction is relatively low for extragalactic objects, at least compared to the optical extinction, thus, it can be ignored for the NIRCam bands. On the other hand, for $z > 2$ SNe, the observer-frame NIR corresponds to rest-frame optical bands, thus, extinction within the host galaxy may be significant. Since the $c$ parameter contains a mixture of both the SN intrinsic color and the reddening due to dust extinction, it is not easy to decide whether we see a peculiar SN Ia that is redder than the normal ones, or whether it is a normal but heavily reddened SN Ia. 

Even though the intrinsic color and dust reddening are highly correlated, it might be possible to disentangle the two effects if one takes into account the expected distribution of ``normal'' SNe Ia on the color-stretch ($c$-$x_1$) diagram. If one allows host galaxy extinction added to the the SALT3-NIR templates, then, for a SN Ia that appears red, the resulting best-fit SALT3 parameters could be closer to the distribution of ``normal'' SNe Ia on the color-stretch diagram than without allowing host extinction to be taken into account. In such a case, it is more probable that the extreme red color is due to host extinction rather than the peculiarity of the SN. 

Therefore, in this study we added host galaxy extinction with the usual $E(B-V)_{\rm host}$ parameter to modify the SALT3-NIR templates in the rest frame. This is a slightly modified model compared to that of P24, who mixed the effect of host extinction with the intrinsic SN color and applied a single $c$ parameter in their fitting. 

We computed  the synthetic light curves in the NIRCam bands from the SALT3-NIR templates within the {\tt sncosmo}\footnote{\url{https://github.com/sncosmo/sncosmo}} Python environment \citep{barbary16}. Two slightly different models were implemented to fit the observed data. In Model~A we used $\log_{10}(x_0)$, $x_1$ and $c$ as independent SALT3 fitting parameters (within bounds, see below), and the distance modulus was found via Eq.~(\ref{eq:mu}). In Model~B we applied Eq.~(\ref{eq:Mbpeak}) to constrain $M_B$ via $M_B^0$, $x_1$ and $c$, and calculated the luminosity distance ($D_L$) to the SN directly from redshift via a cosmological model. We adopted a flat $\Lambda$CDM cosmology with $H_0 = 70.0$~km~s$^{-1}$~Mpc$^{-1}$ and $\Omega_{\rm M} = 0.315$, which is the same model used by P24. Thus, in Model~B the SALT3 fitting parameters are only $x_1$ and $c$, and $x_0$ is constrained by redshift and the cosmological parameters via $D_L$. In both models the SALT3 fitting parameters were supplemented with $E(B-V)_{\rm host}$ (the reddening caused by dust extinction within the host galaxy) and $dt_{\rm max}$ (the difference between the observed moment of $B$-band maximum and a fixed but arbitrary reference epoch).

The reason for using Model~B, and not Model~A alone, was to test whether the data could be reasonably fit by a SN Ia with ``normal'' SALT3 parameters placed at the distance implied by its redshift in the given cosmological model. If the best-fit SALT3 parameters in Model~B ($x_1$ and $c$) were substantially different from those inferred from Model~A, then it would suggest a correlation of these parameters with $x_0$, which would weaken any conclusion implied by the SALT3 parameters for the normal or peculiar nature of SN~2023adsy. On the other hand, the overall agreement between the SALT3 parameters from Model~A and B would strengthen such a conclusion.

To compute the SALT3-NIR fitting via $\chi^2$ minimization, we applied the Price algorithm \citep{brachetti97, manos13}, a controlled random-search technique that uses $N>100$ realizations of the model, each with different parameter values (within their pre-defined bounds), and shifts them toward the location of the absolute minimum within the bounded volume of the $\chi^2$ hyperspace via a Monte-Carlo iteration process. We used $\Delta \chi^2 / \chi^2_{\rm min} \leq 0.1$ as the stopping criterion for the iteration, where $\Delta \chi^2 = \chi^2_{\rm max} - \chi^2_{\rm min}$ for the $N$ different model realizations. Since the geometry of the $\chi^2$ hypersurface can be quite complex, we defined the best-fit value of each parameter as the median of that particular parameter of the $N$ random models after they reached the stopping criterion, instead of exclusively using the model with the lowest $\chi^2$. The distribution of the same random models around the minimum were used to estimate the uncertainty of each best-fit parameter value. 

\begin{table*}[]
    \centering
    \caption{Best-fit parameters from the SALT3-NIR templates. }
    \small
    \begin{tabular}{llcccccccc}
        \hline
        \hline
        $dt_{\rm max}$ & $\log_{10}(x_0)$ & $x_1$ & $c$ & $E(B-V)_{\rm host}$ & $m_B$ & $M_B$ & $\mu$ & $\chi^2$ & ref. \\
        (day) & & & & (mag) & (mag) & (mag) & (mag) & & \\  
        \hline
        8.81(0.73) & -7.04 (0.19) & 2.11 (0.44) & 0.30 (0.13) & 0.68 (0.11) & 28.20 (0.48) & -18.59 (0.40) & 46.79 (0.22) & 1.9249 & Model A \\
        8.82 (0.66) & -7.29 (---) & 2.39 (0.31) & 0.47 (0.08) & 0.54 (0.06) & 28.83 (0.26) & -18.16 (0.26) & 47.00 (0.26) & 1.9543 & Model B \\
        -0.02 (1.60) & -8.15 (0.05) & -0.11 (1.04) & 0.92 (0.05) & 0.00 (---) & 30.73 (0.15) & -16.45 (0.32) & 47.18 (0.28) & 6.4655 & P24 \\
        \hline
    \end{tabular}
    \tablefoot{Uncertainties are shown in parentheses.}
    \label{tab:salt3}
\end{table*}

\begin{table*}[]
    \centering
    \caption{Same as Table~\ref{tab:salt3} but for BayeSN.}
    \small
    \begin{tabular}{lccccccc}
        \hline
        \hline
        $dt_{\rm max}$\tablefootmark{a} & $A_V$ & $\theta_1$ & $\mu_{\rm BSN}$ & $R_V$ & $E(B-V)$ & $\mu$ & Ref.\\
        (day) & (mag) & & (mag) & (fixed) & (mag) & (mag) & \\
        \hline
        4.61 (2.54) & 3.02 (0.15) & -0.97 (0.66) & 46.35 (0.11) & 2.89 & 1.04 (0.05) & 46.77 (0.43) & M20 \\
        5.47 (2.54) & 3.08 (0.15) & -1.05 (0.66) & 46.31 (0.11) & 3.10 & 0.99 (0.05) & 46.73 (0.43) & M20 \\
        0.27 (3.48) & 2.17 (0.25) & 0.33 (0.77) & 46.90 (0.18) & 2.61 & 0.83 (0.10) & 47.33 (0.45) & T21 \\
        0.31 (3.36) & 2.40 (0.27) & 0.37 (0.75) & 46.66 (0.19) & 3.10 & 0.77 (0.09) & 47.08 (0.45) & T21 \\
        4.69 (2.70) & 2.90 (0.17) & -0.91 (0.77) & 46.49 (0.10) & 2.66 & 1.09 (0.06) & 46.91 (0.42) & W22 \\
        7.27 (2.66) & 3.04 (0.17) & -0.99 (0.76) & 46.37 (0.11) & 3.10 & 0.98 (0.05) & 46.79 (0.43) & W22 \\
        \hline
    \end{tabular}
    \tablefoottext{a}{Converted to observer's frame}
    \label{tab:bayesn}
\end{table*}

The parameter bounds, which define the search volume in the parameter space, were defined in a similar way to P24: $\log_{10}(x_0) = [-9,-5]$, $x_1 = [-3,3]$, $c = [-1.5,1.5]$, $E(B-V)_{\rm host} = [0,2]$, $dt_{\rm max} = [-10,10]$. We used flat priors for the distribution of these parameters, and no dependence on each other was assumed, except for the constraint of $x_0$ via $x_1$ and $c$ in Model B (see above). The reference epoch was set to $t_0 = 60240.0$, which is close to the best-fit peak time found by P24.

\subsection{Fitting with BayeSN}\label{sec:bayesn}

BayeSN \citep{mandel22} is another SED-based light-curve fitter for SNe Ia, which extends from the optical to NIR bands and is implemented in a hierarchical Bayesian framework. Being a continuous SED-model, it has the advantage of eliminating the need to use only standard filter sets and precomputed $K$ corrections for each filter, thus, it can be robustly applied to data taken with other filters, provided the new filters are within the wavelength range of the training sets. We utilized the GPU-accelerated python implementation\footnote{https://github.com/bayesn/bayesn} given by \citet{grayling24}, which contains three built-in SED models: the M20 model \citep{mandel22}, the T21 model \citep{thorp21}, and the W22 model \citep{ward23}. These models have been trained on different data and filter sets: M20 is trained on the data given by \citet{avelino19} using BVRIYJH filters (wavelength range 3000 - 18500 \AA), T21 is trained on Foundation DR1 data \citep{foley18, jones19} using griz filters (wavelength range 3500 - 9500 \AA), and the W22 model is trained on the combination of the previous two data and filter sets. The Milky Way extinction map used for all three models is the one by \citet{sf11}. More technical details can be found in the BayeSN documentation online\footnote{https://bayesn.readthedocs.io/en/latest/}. 

Since the JWST NIRCam filter sets are not included in the BayeSN implementation, those were added to the BayeSN database manually using the filter transmission curves downloaded from the JWST User Documentation website\footnote{https://jwst-docs.stsci.edu/jwst-near-infrared-camera/nircam-instrumentation/nircam-filters}. 
BayeSN has a built-in Markov Chain Monte Carlo (MCMC) fitting routine to fit individual light curves with any of its pretrained SED models. The fitting parameters are $dt_{max}$ (same as for SALT3, but in rest frame instead of observer frame), $A_V$ (total line-of-sight V-band extinction in magnitudes), $\theta_1$ (the light-curve parameter, analogous to the stretch parameter $x_1$ in SALT3), $\mu$ (extinction-free distance modulus in magnitudes), and $R_V$ ($=A_V / E(B-V)$, the reddening-law parameter). Note that during the training the same $R_V$ was assumed for each SNe in the training set, thus the default $R_V$ for each model is a mean value obtained from the corresponding training set. In principle, $R_V$, can be treated as a free fitting parameter if BayeSN is applied to suitable input datasets. Since the poorly sampled light curve of SN~2023adsy does not cover a suitably wide rest-frame wavelength regime, we used BayeSN with $R_V$ values fixed at either its default (trained) value or $R_V=3.1$, which corresponds to the Milky Way extinction law.     

The BayeSN distance scale depends on the distance moduli of the SNe in the training samples. \cite{mandel22} explore the consistency between the BayeSN and SALT2 distance scales by using the Tripp-relation for the training sample with the following parameters:
\begin{equation}
    \mu_{\rm ext} = m_B + 19.01 + 0.117 \cdot x_1 - 2.939 \cdot c, 
    \label{eq:muext}
\end{equation}
which is somewhat different from the parametrization we used for computing SALT3-NIR distances (see Eqs.~\ref{eq:Mbpeak} and \ref{eq:mu}). Thus, in order to convert the BayeSN distance moduli to the distance scale we adopted for SALT3-NIR, we find 
\begin{equation}
    \mu_{\rm ext} = 1.017 (\pm 0.012) \cdot \mu_{\rm BSN} - 0.596 (\pm 0.413),
    \label{eq:mubsn}
\end{equation}
where $\mu_{\rm ext}$, as above, is the distance modulus for a SN Ia in the training (``external'') sample and $\mu_{\rm BSN}$ is the distance modulus predicted by BayeSN (see Table~1 in \citealt{mandel22}). Then, after applying the Tripp-relation with the parameters adopted in Sect.~\ref{sec:salt3-nir} to the training sample of \citet{mandel22}, we get
\begin{equation}
    \mu = 1.0031 (\pm 0.0005) \cdot \mu_{\rm ext} + 0.082 (\pm 0.018),
    \label{eq:mur22}
\end{equation}
where $\mu$ is the distance modulus compatible with the Pantheon+SH0ES \citep{riess22} distance scale. 

\section{Results and discussion} \label{sec:results}

This section contains the results of fitting the NIRCam photometry of SN~2023adsy with SALT3-NIR and BayeSN. It also contains a discussion of these results.

\begin{figure*}
    \centering
    \includegraphics[width=0.49\linewidth]{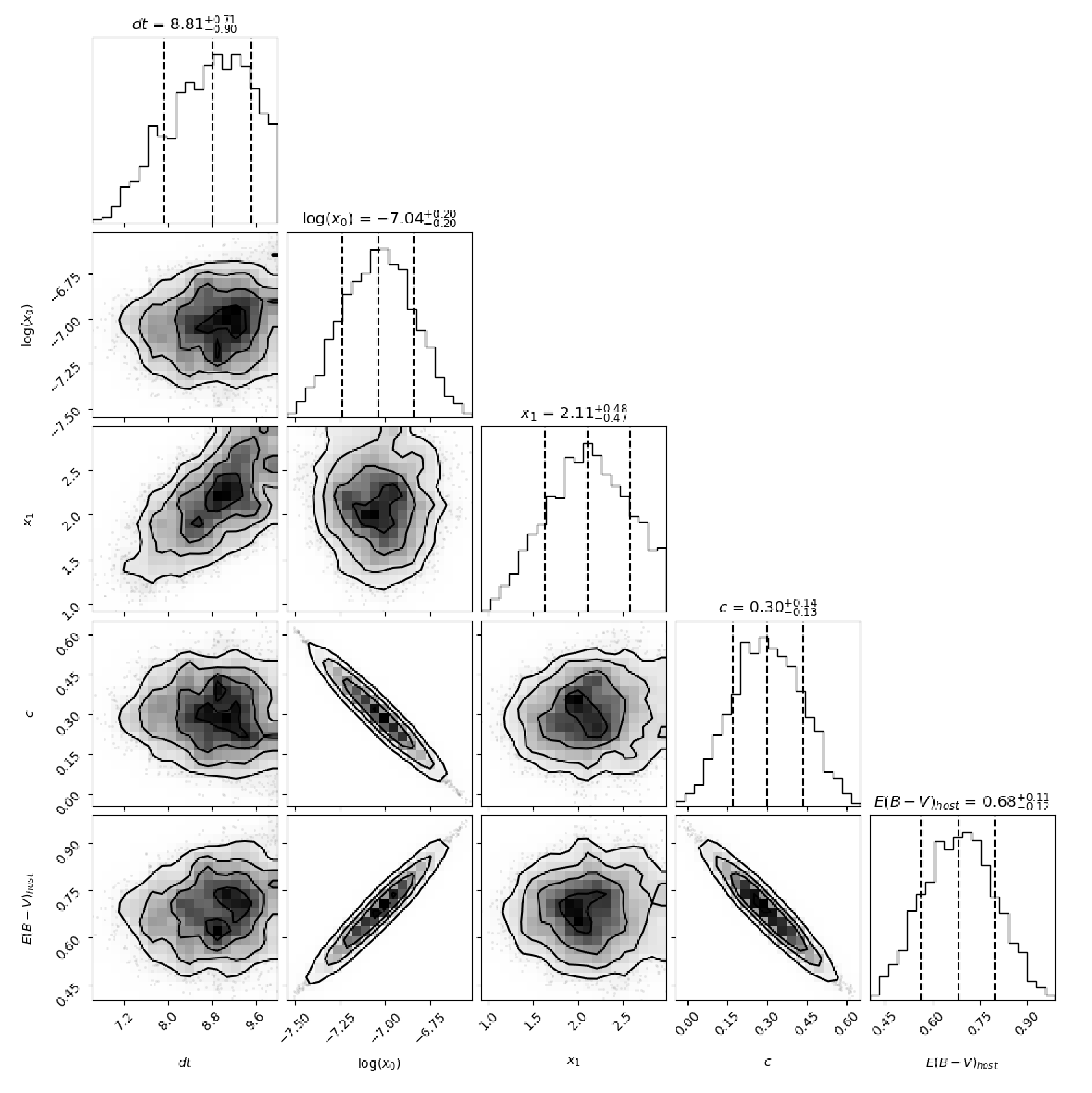}
        \includegraphics[width=0.49\linewidth]{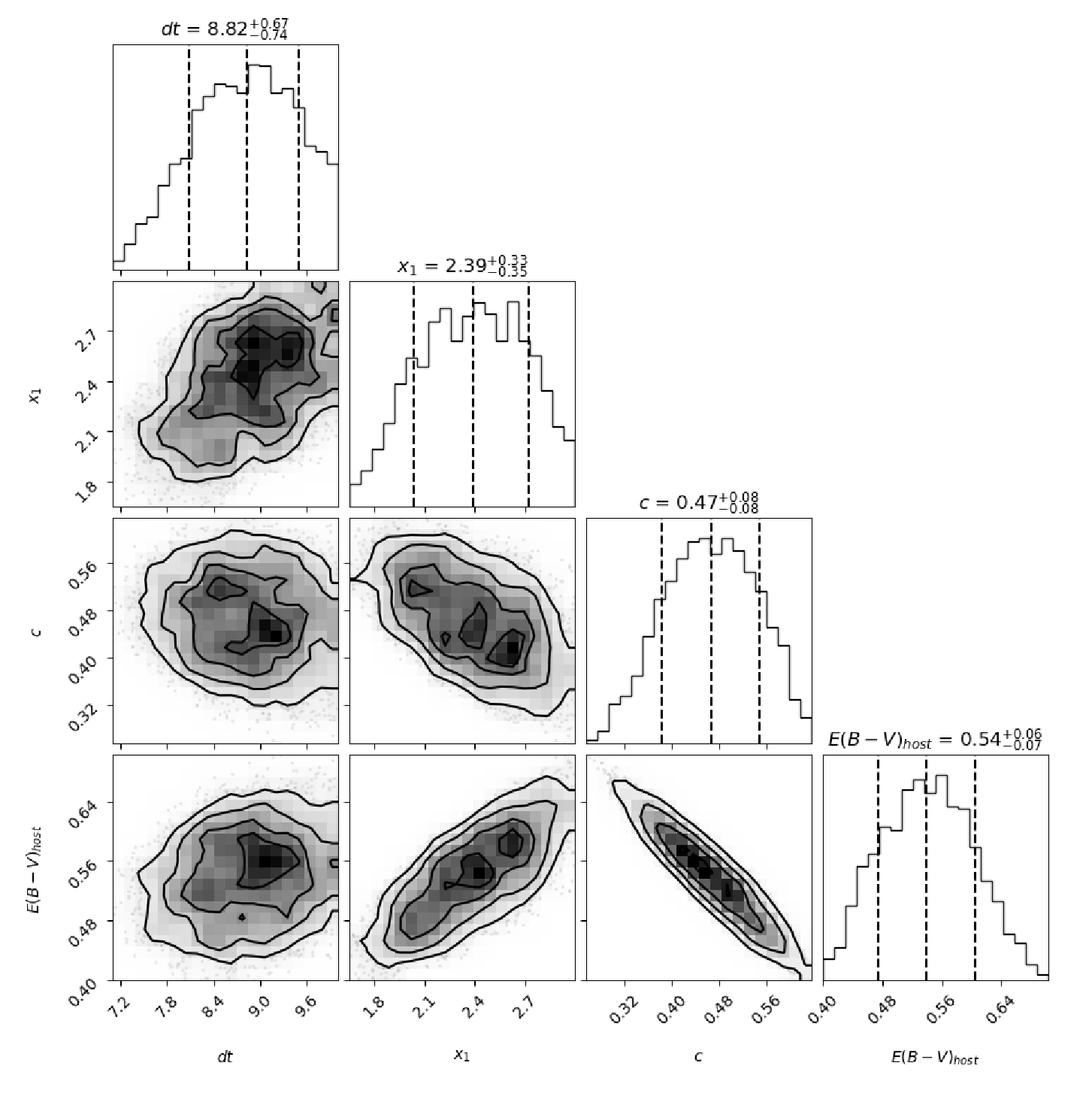}
    \caption{Corner plots showing the distribution of fitting parameters for Model~A (left panel) and Model~B (right panel). Contours show the 1-, 2-, and 3-$\sigma$ levels that correspond to quantiles $q=0.16, 0.50,$ and $0.84$, respectively.}
    \label{fig:corner_x0}
\end{figure*}

\begin{figure*}
\centering
\includegraphics[width=0.48\linewidth]{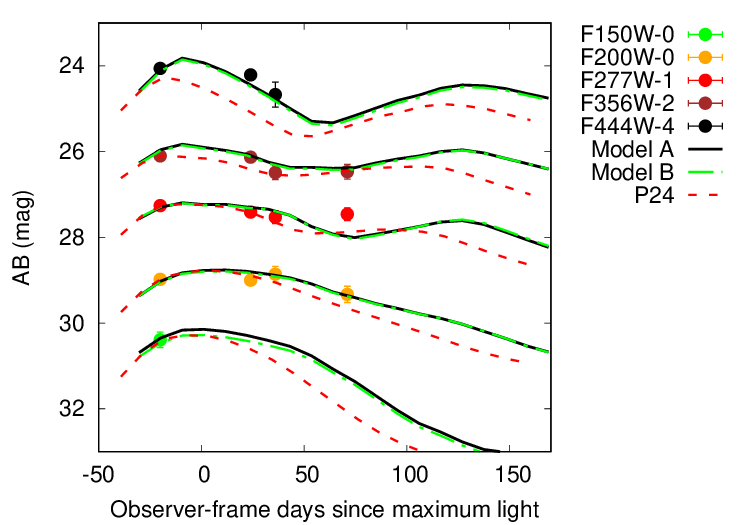}
\includegraphics[width=0.48\linewidth]{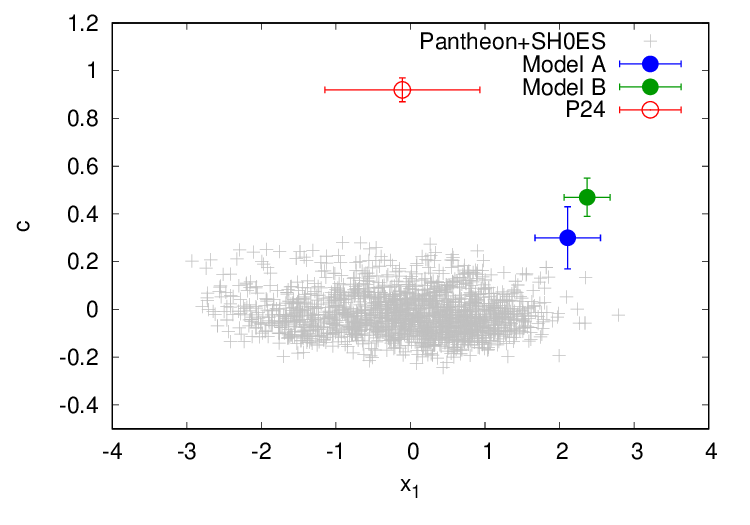}
\caption{Left panel: NIRCam photometric data for SN~2023adsy (circles) and the best-fit SALT3-NIR templates (curves). Solid curves: Model A; dashed curves: Model B; dash-dotted curves: P24. The applied vertical shifts (in magnitudes) for better visibility are indicated in the legends.
Right panel: Best-fit SALT parameters for SN~2023adsy (filled circles) compared to those from P24 (open circle) and other normal type Ia SNe from the Pantheon+ sample (gray plus signs). }
\label{fig:salt3fit}    
\end{figure*}

\begin{figure}
    \centering
    \includegraphics[width=1.0\linewidth]{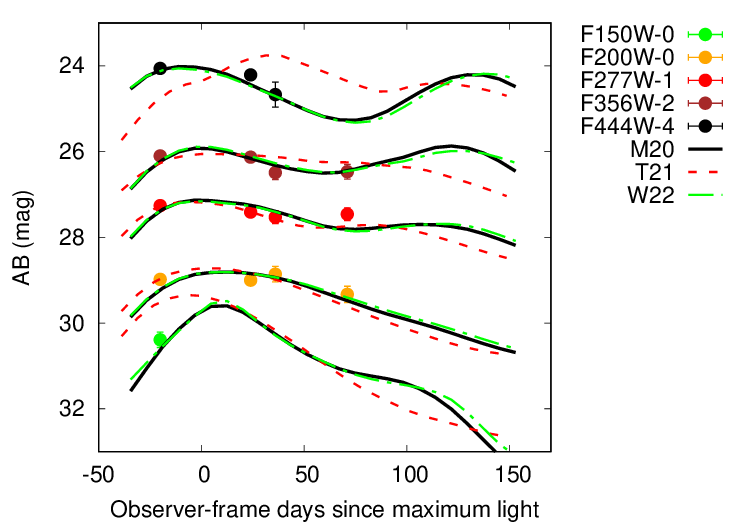}
    \caption{Fitting of SN 2023adsy NIRCam data with BayeSN templates using $R_V=3.1$. }
    \label{fig:bayesnfit}
\end{figure}

\begin{figure*}
    \centering
    \includegraphics[width=0.8\linewidth]{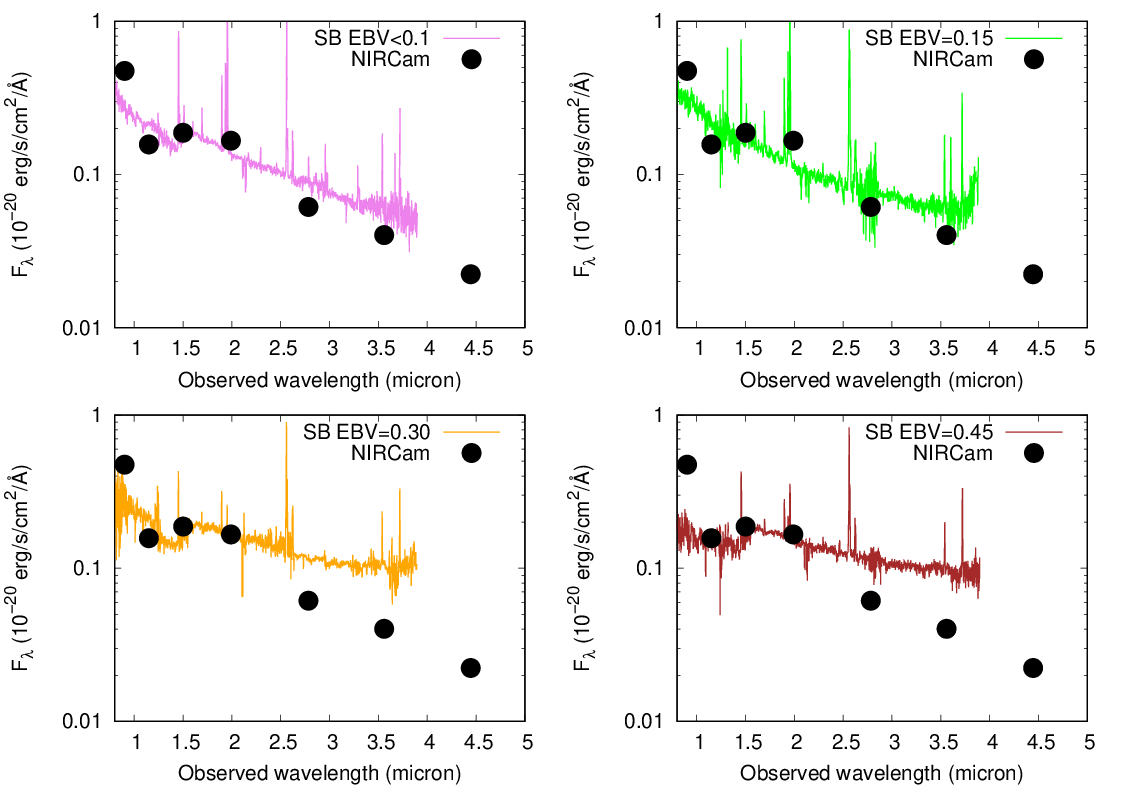}
    \caption{NIRCam photometry of the host galaxy taken before the SN discovery (black circles) compared to templates of star-forming galaxies with different overall dust attenuation (colored curves). }
    \label{fig:host}
\end{figure*}

\subsection{SALT3-NIR fitting}

The best-fit parameters of Model~A and B, as well as the solution found by P24, are collected in Table~\ref{tab:salt3}. The uncertainties, inferred as the standard deviation of the distribution of the feasible models (see Sect.~\ref{sec:model}), are given in parentheses. 

Corner plots for the best-fit parameters of Model~A and B are shown in Fig.~\ref{fig:corner_x0}.
As expected, in Model~A the flux-scaling factor ($x_0$) strongly correlates with both the color term ($c$) and the host reddening ($E(B-V)_{\rm host}$. This results in an especially wide distribution of the possible values of the color term: the feasible models with $\chi^2 \leq 1.1 \chi^2_{\rm min}$ have their color term between $0.1 \lesssim c \lesssim 0.5$.    

The resulting light curves from Model~A (solid curves) and Model~B (dashed curves) together with the NIRCam photometry of SN~2023adsy (filled circles) are shown in the left panel of Fig.~\ref{fig:salt3fit}. For comparison, we also plot the best-fit light curves of P24 (dash-dotted curves).  We see that the differences between the light curves from Model A and B are negligibly small. The differences between our best-fit light curves and those of P24 are also relatively minor, and all model light curves fit the observed data quite well. This is also reflected by the low $\chi^2$ values of the best-fit solutions shown in Table~\ref{tab:salt3}. The slightly higher $\chi^2$ of the P24 solution is caused by the F444W filter data, which could be fit somewhat worse with their model compared to ours. Note that for the same reason P24 discarded the F444W data from their fitting. 

\begin{figure}
    \centering
    \includegraphics[width=1.0\linewidth]{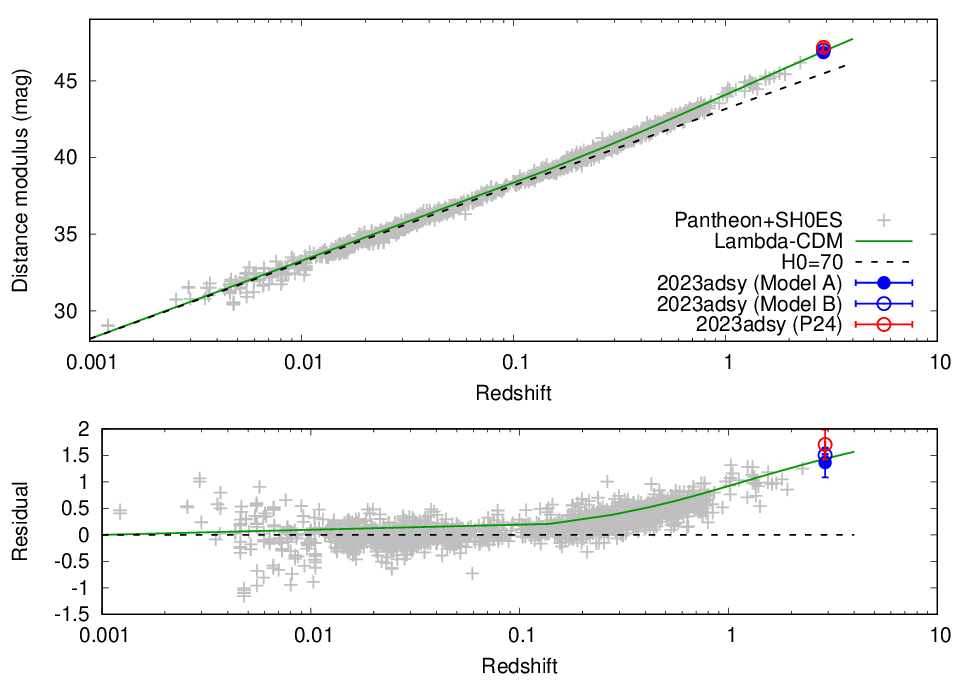}
    \caption{Hubble diagram for SNe Ia, including SN~2023adsy. The green line indicates the flat $\Lambda$CDM model with $\Omega_M = 0.3$.}
    \label{fig:hubble}
\end{figure}

In the right panel of Fig.~\ref{fig:salt3fit} the location of the best-fit parameters are shown on the $x_1 - c$ plane for each model. Our Model~A and B are plotted with filled circles, while the solution found by P24 is represented by the open circle. The gray symbols correspond to the ``normal'' SNe Ia in the Pantheon+ sample \citep{scolnic22}.  It is clear that SN~2023adsy is somewhat off from the distribution of the bulk of SNe Ia on this diagram, but our best-fit solutions (Model A and B) are less extreme than the result by P24.  For Model A, the best-fit stretch and color ($x_1$ and $c$) parameters are near the edge of the distribution of the SNe Ia in the Pantheon+ sample. The result from Model B is similar, but slightly more stretched and redder compared to the result from Model A. 

In contrast, the solution by P24 represents a less stretched (close to $x_1 \sim 0$), but very red ($c \sim 0.9$) SN Ia. Such a SN Ia, with a very red color but a normal decline rate, would be quite unusual, at least among the 
low-z SN Ia sample. P24 also pointed out that their result, if true, suggests that SN~2023adsy would be peculiar: it is a normal Ia according to the light-curve decline rate, a 91bg-like Ia according to its B-band absolute magnitude, but similar to a Ca-rich transient according to its very red color and Ca~II velocity, even though it is $\sim 1$ mag brighter at peak than most Ca-rich transients. 

The new solutions presented in this paper (both Model A and B) suggest that this is not necessarily the case. Instead of being a peculiar object with ``mixed'' properties of low-z transients, SN~2023adsy could be more similar to normal low-z SNe Ia, even though it is near the edge of their parameter distribution. Based on both its new best-fit stretch and color (see Table~\ref{tab:salt3}), SN~2023adsy would  pass the selection criteria for being a ``normal'' SN Ia that can be used e.g. for cosmological analysis, if it was a low-z object \citep{scolnic18}. Note that the new stretch parameter, which is definitely high but not extreme, might still be somewhat overestimated, because, as seen in the left panel of Fig.~\ref{fig:salt3fit}, the NIRCam photometry does not cover the second peak in the reddest filters, which would be the best constraint for the stretch parameter. Still, SN~2023adsy could be a ``normal'' SN Ia at $z\sim 2.9$, instead of being a peculiar object suggested by the results of P24, if most of its extremely red color is due to dust extinction in its host galaxy.

\subsection{BayeSN fitting}

Figure~\ref{fig:bayesnfit} shows the best-fit light curves from fitting the M20, T21, and W22 pretrained models by BayeSN to the NIRCam data. Note that this figure only shows the results that correspond to $R_V=3.1$, because the other model fits look practically the same, the differences are tiny. We see that the M20 and W22 models give consistent fits to the data, while the best-fit light curves from the T21 model are inferior with respect to the other two solutions. This is probably due to the fact that the T21 SEDs were trained on a much narrower wavelength range than the M20 and W22 SEDs. Indeed, in Fig.~\ref{fig:bayesnfit} we see that the T21 fits are the worst for the data in the F150W and F444W filters, corresponding to $\sim 0.38$ micron and $1.13$ micron rest-frame wavelengths at $z=2.9$. These wavelengths are outside the wavelength range of the T21 training sample (see Sect.~\ref{sec:bayesn}), therefore BayeSN is not expected to give a good prediction for these data.   

The parameters for the best-fit BayeSN models are collected in Table~\ref{tab:bayesn}. Here, to make the comparison with the SALT3-NIR fitting results easier, we also show the corrected $\mu$ distance modulus (via Eqs. \ref{eq:mubsn} and \ref{eq:mur22}) as well as the $E(B-V)$ reddening derived from $A_V$ and $R_V$. 

BayeSN also predicts high reddening toward SN~2023adsy: the $E(B-V)$ values are almost twice as high as the best-fit $E(B-V)_{\rm host}$ values from SALT3-NIR. This is not surprising, because BayeSN models red SNe Ia entirely by assuming dust extinction, unlike SALT3-NIR that takes into account both dust extinction and intrinsic color. Thus, BayeSN assumes that the red color of SN~2023adsy is due to dust extinction alone, while in our SALT3-NIR model it is because SN~2023adsy is both intrinsically red and suffers from considerable dust extinction within its host. This explains why BayeSN found almost a factor of two higher  $E(B-V)   $. 

The other parameter that is common in the two codes is $\mu$, the distance modulus. After correcting the $\mu_{\rm BSN}$ distance moduli to the Pantheon+SH0ES distance scale (Eqs. \ref{eq:mubsn} and \ref{eq:mur22}), the results are entirely consistent with the distance moduli inferred from SALT3-NIR. Note that although the BayeSN best-fit distance moduli ($\mu_{\rm BSN}$) have lower uncertainties than those from SALT3-NIR, the corrected $\mu$ values are more uncertain due to the $\mu_{\rm BSN} \rightarrow \mu$ transformation (caused mostly by the zero point uncertainty in Eq.~\ref{eq:mubsn}). Despite the higher uncertainty, we conclude that the BayeSN distances are in very good agreement with the results of SALT3-NIR.   

Regarding the normal or peculiar nature of SN~2023adsy, the BayeSN parameter $\theta_1$ plays a similar role to the stretch parameter $x_1$ in SALT3: $\theta_1$ is roughly inversely proportional to $x_1$ and $\theta_1 = 0$ corresponds to $x_1 \approx -0.8 \pm 0.2$ \citep{mandel22}. From Table~\ref{tab:bayesn}, $\theta_1 \sim -1 \pm 0.1$ for the preferred M20 and W22 models, which, if taken at face value, would imply $x_1 \sim 0.25 \pm 0.25$ based on Fig.~4 of \citet{mandel22}. Thus, BayeSN also finds that the NIRCam data of SN~2023adsy can be modeled with the SED of a ``normal'' SN Ia, which is consistent with the prediction by SALT3-NIR Model~A and B.

\subsection{The host galaxy}

The NIRCam photometry of the host galaxy, JADES-GS+53.13485$-$27.82088 (JADES Host ID 96906), was downloaded from the JADES archive\footnote{\tt https://archive.stsci.edu/hlsp/jades} \citep{jades23} and converted to $F_\lambda$ fluxes. We compare these observer-frame data with redshifted galaxy templates taken from \citet{kinney96} in Fig.~\ref{fig:host}. We see that the rising SED toward shorter wavelengths strongly suggests a star-forming galaxy with low observable reddening, $E(B-V)_{\rm obs} \lesssim 0.15$ mag, which is similar to the conclusion of P24. This low reddening value, however, does not necessarily imply that the total dust content of the host is low. In fact, significant dust content is more probable in a star-forming galaxy than the absence of dust. The blue observed SED could be the result of the orientation of the host galaxy, being face-on rather than edge-on. If the host is visible face-on, it is possible that we observe many young, blue stars in the volume facing toward the Earth, but the inner part of the galaxy could still suffer from significant dust obscuration and extinction \citep[see e.g.,][]{bowler22}. 
We conclude that the NIRCam photometry of the host galaxy does not rule out high dust extinction within the obscured part of the host, thus, the photometry of SN~2023adsy could be affected by dust extinction despite the blue observed color of its host. 

\subsection{SN~2023adsy on the Hubble diagram}

The position of SN~2023adsy on the Hubble diagram is presented in Fig.~\ref{fig:hubble}, together with the SNe Ia from the Pantheon+SH0ES sample \citep{scolnic22, riess22}. The prediction from the adopted $\Lambda$CDM cosmology (Sect.~\ref{sec:model}) and a simple linear Hubble law with $H_0 = 70$ kms$^{-1}$Mpc$^{-1}$ are also shown as solid and dashed lines, respectively. We see that the inferred distance modulus from both Model A and B is consistent with the prediction of the adopted $\Lambda$CDM model ($\sim 46.98$ mag), just like the best-fit solution of P24, at the $1 \sigma$ level. 

The detection of a single SN Ia at $z>2$ is not sufficient to constrain the physics of the Ia population at high redshifts, but the agreement between the inferred peak luminosity of SN~2023adsy and the prediction of the $\Lambda$CDM cosmology fitted to low-redshift SNe Ia suggests no significant evolution of the physics of SNe Ia up to $z \sim 3$. 

\subsection{Constraints on the SN Ia rates and delay-time distributions at high-z}

\begin{figure*}
    \centering
    \includegraphics[width=0.45\linewidth]{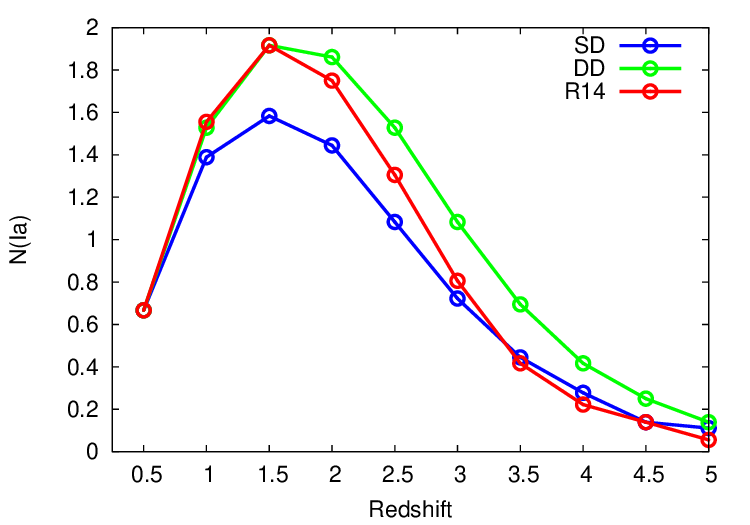}
    \includegraphics[width=0.45\linewidth]{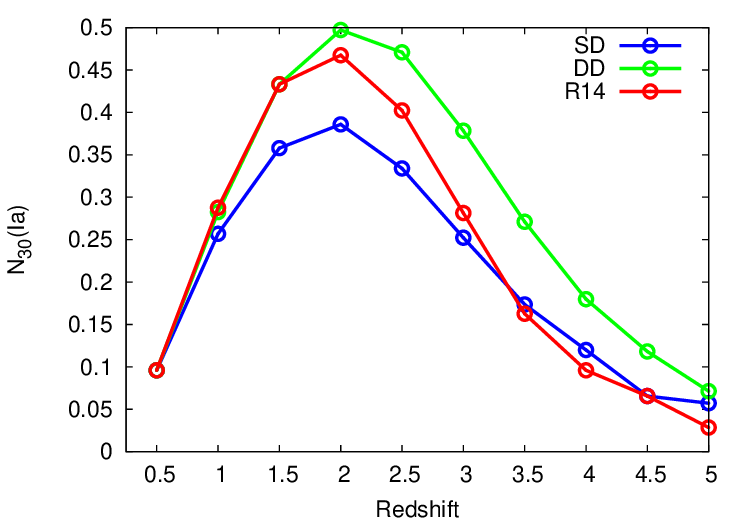}
    \caption{Expected number of SNe Ia as a function of redshift in the JADES survey area ($\sim 25$ arcmin$^2$) in one year (left panel). Different DTD scenarios are shown with different colors as indicated in the legend: SD = single degenerate, DD = double degenerate, R14 = \citet{rodney14}. 
    The right panel shows the expected numbers assuming $\sim 30$ days rest-frame visibility time for a SN Ia. }
    \label{fig:rates}
\end{figure*}

\begin{table*}[]
    \centering
    \caption{Expected numbers of SNe Ia within the JADES survey area in one year. }
    \begin{tabular}{lccccccc}
    \hline
    \hline
    z & z+dz & N(SD) & N$_{30}$(SD) & N(DD) & N$_{30}$(DD) & N(R14) & N$_{30}$(R14) \\
    \hline
    0.5 & 1.0 & 0.666 &  0.096 & 0.666 &  0.096 &  0.666 & 0.096 \\
    1.0 & 1.5 & 1.389 & 0.257 & 1.528 & 0.282 & 1.555 & 0.287 \\
    1.5 & 2.0 & 1.583 & 0.358 & 1.917 & 0.433 & 1.917 & 0.433 \\
    2.0 & 2.5 & 1.444 & 0.386 & 1.861 & 0.497 & 1.750 & 0.467 \\
    2.5 & 3.0 & 1.083 & 0.334 & 1.528 & 0.471 & 1.305 & 0.402 \\
    3.0 & 3.5 & 0.722 & 0.252 & 1.083 & 0.378 & 0.805 & 0.281 \\
    3.5 & 4.0 & 0.444 & 0.173 & 0.694 & 0.271 & 0.417 & 0.163 \\
    4.0 & 4.5 & 0.278 & 0.120 & 0.417 & 0.180 & 0.222 & 0.096 \\
    4.5 & 5.0 & 0.134 & 0.066 & 0.250 & 0.118 & 0.139 & 0.065 \\
    5.0 & 5.5 & 0.111 & 0.057 & 0.139 & 0.071 & 0.055 & 0.028 \\
    \hline
    \end{tabular}
    \tablefoot{See text for explanation.}
    \label{tab:rates}
\end{table*}

Following \citet{rv19}, we estimate the expected numbers of SNe Ia within the JADES survey area as a function of redshift, and compare these predictions with the discovery of a single event, SN~2023adsy. 
The expected number of SNe Ia at a given redshift depends on two quantities: the cosmic SFR and the delay-time distribution (DTD), both as a function of redshift. \citet{rv19} consider a variety of proposed DTD functions and compute different predictions for the numbers of SNe Ia at $z > 1$. Here we use their Table~2, which shows the numbers for the single-degenerate (SD) and double-degenerate (DD) scenarios, as well as the mixed ``prompt+DD'' scenario by \citet{rodney14}. Since those numbers correspond to a different survey area ($300$ arcmin$^2$ instead of 25) and a three-year-long survey time, we rescaled them to the area of the JADES survey and one-year survey time. The results are shown in Table~\ref{tab:rates} and plotted in Fig.~\ref{fig:rates}. The right panel displays the numbers that were inferred after assuming a $\sim 30$ day-long visibility time in the rest frame for each event, which are given in the $N_{30}$ columns in Table~\ref{tab:rates}. 

The numbers in Table~\ref{tab:rates} suggest that within the redshift interval of $1 < z < 2$ the number of SNe Ia that are detectable within the JADES survey area during one year are 3.0, 3.4, and 3.5 for the SD, DD, and R14 scenarios, respectively, while between $2 < z < 3$ these are 2.5, 3.4, and 3.0. Since in these redshift intervals SNe Ia are not expected to be visible for one year, a more realistic estimate could be obtained from the $N_{30}$ numbers that take into account the limited ($\sim 30$ days in rest-frame) visibility of a SN Ia at $z> 1$. Using these estimates, one can get 0.6, 0.7, and 0.7 SNe per year for $1 < z < 2$ and 0.7, 1.0, and 0.9 SNe per year for $2 < z < 3$. The latter numbers are all consistent with the discovery of only one SN Ia, SN~2023adsy at $z \sim 2.9$. The lack of detection of a similar event at $z < 2$ may simply be due to the lower number of potential host galaxies in this redshift interval within the JADES survey area. 

Unfortunately, from a single SN Ia it is not possible to get a realistic constraint on the different DTD scenarios. Still, the discovery of SN~2023adsy with JWST demonstrated the possbility of detecting type Ia SNe at significantly higher redshifts than before, which could lead to further unprecedented discoveries regarding the progenitors of SNe Ia as well as their cosmological implications in the near future. 

\section{Summary} \label{sec:summary}

We re-fit the NIRCam photometry of SN~2023adsy, a type Ia SN discovered by JWST at $z \sim 2.9$ \citep{pierel24}, with SALT3-NIR and BayeSN templates, taking into account a potential extinction due to dust in the host galaxy. We found that SN~2023adsy can be adequately fitted ($\chi^2 / {\rm dof} \sim 2$) with the SALT3-NIR model if $E(B-V)_{\rm host} \lesssim 0.7 \pm 0.1$ mag reddening due to dust is added. Similarly, BayeSN predicts $E(B-V) \sim 1.0 \pm 0.1$ mag. This resulted in a slowly declining ($x_1 \gtrsim 2.1 \pm 0.4$) and moderately red ($c \gtrsim 0.3 \pm 0.1$) SN Ia, which is still within the parameter distribution of ``normal'' SNe Ia, as shown by the Pantheon+ sample. 

Our result implies that SN~2023adsy may not be such an extremely red and faint but slowly declining and high-velocity SN Ia as found by P24, which, if true, would be a very peculiar object. The present result suggests that SNe Ia at $z\lesssim 3$ could be more or less similar to their local counterparts, and their light curves can be standardized using the same techniques that work for local SNe Ia. 

The Ia discovery rate from the JADES survey, one confirmed SN Ia in $2 < z < 3$ per year, is consistent with the expected SN Ia rates at this redshift range. Even though a single event cannot constrain the different DTD models, more discoveries, like SN~2023aeax \citep{pierel24b}, can significantly improve our understanding of the physics of SNe Ia.

\begin{acknowledgements}
This research is supported by NKFIH-OTKA grant K142534 from the National Research, Development and Innovation Office, Hungary.  The SN group at Konkoly Observatory was supported by the project ``Transient Astrophysical Objects'' GINOP 2.3.2-15-2016-00033 by the Hungarian Government, funded by the European Union. 
\end{acknowledgements}

\bibliographystyle{aa}
\bibliography{aa54209-25}

\end{document}